\documentclass[graybox]{svmult}
\usepackage{hyperref}
\usepackage{mathptmx}       
\usepackage{helvet}         
\usepackage{courier}        
\usepackage{makeidx}         
\usepackage{graphicx}        
\usepackage{tensor}
\usepackage{multicol}        
\usepackage[bottom]{footmisc}
\usepackage{amsmath,amssymb,amsfonts}        
\makeindex             
\usepackage{enumitem}
\usepackage{tensor}
\usepackage{physics}

\newcommand{\overbar}[1]{\mkern 1.5mu\overline{\mkern-1.5mu#1\mkern-1.5mu}\mkern 1.5mu}

\def\TT{{T\overbar{T}}}

\usepackage{soul}

\begin{document}
\title*{Field-Dependent Metrics and Higher-Form Symmetries in Duality-Invariant Theories of Non-Linear Electrodynamics}
\titlerunning{Field-Dependent Metrics, Higher-Form Symmetries, Duality-Invariant Electrodynamics}
\author{Christian Ferko and Cian Luke Martin}
\institute{Christian Ferko \at University of California, Davis, 1 Shields Avenue, Davis, CA 95616, USA. \\ \email{caferko@ucdavis.edu}
\and Cian Luke Martin \at University of Queensland, St Lucia QLD 4072, Australia. \\ \email{c.lukemartin@uq.edu.au}
}
%
%
\maketitle

\vspace{-9pt}

\abstract{We prove that a $4d$ theory of non-linear electrodynamics has equations of motion which are equivalent to those of the Maxwell theory in curved spacetime, but with the usual metric $g_{\mu \nu}$ replaced by a unit-determinant metric $h_{\mu \nu} ( F )$ which is a function of the field strength $F_{\mu \nu}$, if and only if the theory enjoys electric-magnetic duality invariance. Among duality-invariant models, the Modified Maxwell (ModMax) theory is special because the associated metric $h_{\mu \nu} ( F )$ produces identical equations of motion when it is coupled to the Maxwell theory via two different prescriptions which we describe. We use the field-dependent metric perspective to analyze the electric and magnetic $1$-form global symmetries in models of self-dual electrodynamics. This analysis suggests that any duality-invariant theory possesses a set of conserved currents $j^\mu$ which are in one-to-one correspondence with $2$-forms that are harmonic with respect to the field-dependent metric $h_{\mu \nu} ( F )$.
}


\section{Introduction}\label{sec:intro}

Exploiting the symmetries of a classical or quantum field theory -- and the consequences of these symmetries, such as conservation laws or Ward identities -- is a well-established tool in studying such a theory's dynamics. Although the power of symmetry has been understood by physicists for a long time, going back at least to the work of Noether and Wigner, recent developments have expanded our notion of what constitutes a symmetry of a physical system. The framework of generalized global symmetries offers a unified perspective in which several different symmetry-like structures can be described using a common language, and has led to many new results such as constraints on renormalization group flows and insights on the infrared phase structure of interacting gauge theories; see \cite{Sharpe:2015mja,Gomes:2023ahz,Schafer-Nameki:2023jdn,Brennan:2023mmt,Bhardwaj:2023kri,Shao:2023gho} for reviews.

Although many sub-classes within the broader umbrella of generalized global symmetries have been studied -- such as non-invertible symmetries, categorical symmetries, and higher-group symmetries -- the first members of this class to receive widespread notice were higher-form global symmetries, which were introduced in \cite{Gaiotto:2014kfa}. Whereas the objects which are charged under a conventional global symmetry of a quantum field theory, like the $U(1)$ symmetry associated with phase rotations of a complex scalar field, are point-like, the operators which are charged under these higher-form global symmetries are extended objects, such as lines and surfaces.

Higher-form symmetries are commonly studied in complete microscopic quantum field theories, which are well-defined at short distances, such as the Maxwell conformal field theory. However, within effective field theories which give an approximate description of dynamics at long distances, there can be ``emergent'' higher-form global symmetries which are nonetheless broken at short distances \cite{Cherman:2023xok}.

In this article, we will be interested in investigating the consequences of such emergent higher-form symmetries in theories of source-free non-linear electrodynamics in four spacetime dimensions. We view these as effective field theories for an Abelian gauge field $A_\mu$ with field strength $F_{\mu \nu}$, described by a Lagrangian
\begin{align}
    \mathcal{L} = \mathcal{L} ( F_{\mu \nu} ) \, ,
\end{align}
which is a function of the field strength but not of its derivatives. Although the most commonly studied member of this family, the Maxwell theory with Lagrangian $\mathcal{L} = - \frac{1}{4} F^{\mu \nu} F_{\mu \nu}$, is free, a general model in this class will exhibit interactions. An example of an interacting model is the Born-Infeld theory, whose Lagrangian is
\begin{align}\label{born_infeld}
    \mathcal{L}_{\text{BI}} = T \left( 1 - \sqrt{ 1 + \frac{1}{2 T} F^{\mu \nu} F_{\mu \nu} - \frac{1}{16 T^2} \left( F^{\mu \nu} \widetilde{F}_{\mu \nu} \right)^2 } \right) \, ,
\end{align}
where $\widetilde{F}^{\mu \nu} = \frac{\sqrt{ - \det ( g ) }}{2} \epsilon^{\mu \nu \rho \sigma} F_{\rho \sigma}$ denotes the Hodge dual of $F_{\mu \nu}$. The Lagrangian (\ref{born_infeld}) describes the effective dynamics of the gauge field on the worldvolume of a D3-brane in string theory, and the parameter $T$ is the tension of this brane. Another interesting, and recently discovered, model of non-linear electrodynamics is the Modified Maxwell or ModMax theory \cite{Bandos:2020jsw}, which is described by the Lagrangian
\begin{align}\label{modmax}
    \mathcal{L}_{\text{ModMax}} = \frac{1}{4} \left( - \cosh ( \gamma ) F^{\mu \nu} F_{\mu \nu} + \sinh ( \gamma ) \sqrt{ \left( F^{\mu \nu} \right)^2 + \left( F^{\mu \nu} \widetilde{F}_{\mu \nu} \right)^2 } \right) \, .
\end{align}
Both the Born-Infeld theory (\ref{born_infeld}) and the ModMax theory (\ref{modmax}) enjoy the additional property of invariance under electric-magnetic duality rotations, much like the ordinary Maxwell theory.\footnote{We will use the terms ``duality-invariant'' and ``self-dual'' interchangeably to refer to any theory of non-linear electrodynamics with this property.}  In fact, the ModMax theory (\ref{modmax}) can be characterized as the unique conformally invariant and electric-magnetic duality-invariant model which is a continuous deformation of the Maxwell theory, under the assumption that no derivatives of the field strength appear in the Lagrangian. For an introduction to these and other theories of non-linear electrodynamics, see \cite{Sorokin:2021tge}.

The primary tool which we will use in our analysis of the generalized global symmetries of theories of self-dual electrodynamics is that of field-dependent metrics. In the past few years, this subject has undergone substantial development \cite{Conti:2018tca,Tolley:2019nmm,Conti:2022egv,Morone:2024ffm,Babaei-Aghbolagh:2024hti,Tsolakidis:2024wut} in connection with classical deformations of Lagrangians that are driven by functions of the energy-momentum tensor. The most famous of these is the $\TT$ deformation of two-dimensional quantum field theories \cite{Smirnov:2016lqw,Cavaglia:2016oda}, which is well-defined at the quantum level due to the properties of the point-split $\TT$ operator \cite{Zamolodchikov:2004ce}. At the classical level, a version of the $\TT$ flow for the Lagrangian can be defined in any spacetime dimension $d$ by the differential equation
\begin{align}\label{TT_flow_eqn}
    \frac{\partial \mathcal{L}^{(\lambda)}}{\partial \lambda} = \frac{1}{2d} \left( T^{(\lambda) \mu \nu} T^{(\lambda)}_{\mu \nu} - \frac{2}{d} \left( \tensor{T}{^{(\lambda)}^\mu_\mu} \right)^2 \right) \, ,
\end{align}
along with an initial condition $\mathcal{L}^{(0)}$. Here $T_{\mu \nu}^{(\lambda)}$ is the conserved energy-momentum tensor (also called the stress tensor) associated with the Lagrangian $\mathcal{L}^{(\lambda)}$. For $d = 2$, it was shown that solutions to the classical equations of motion associated with the Lagrangian $\mathcal{L}^{(\lambda)}$ which solves (\ref{TT_flow_eqn}) can be generated from solutions to the undeformed equations of motion associated with $\mathcal{L}^{(0)}$ by enacting a diffeomorphism which depends on the stress tensor \cite{Conti:2018tca}. This change of coordinates produces a field-dependent metric, which gives a ``geometrization'' of the solution to the flow (\ref{TT_flow_eqn}).

A similar geometrical interpretation of stress tensor flows is available in $d > 2$ dimensions \cite{Conti:2022egv,Morone:2024ffm}. This is directly relevant for our present purposes since several theories of non-linear electrodynamics satisfy such stress tensor flows. For instance, the Born-Infeld theory (\ref{born_infeld}) satisfies the $4d$ version of the $\TT$ flow (\ref{TT_flow_eqn}) with initial condition given by the Maxwell Lagrangian \cite{Conti:2018jho}. Likewise, the ModMax Lagrangian (\ref{modmax}) satisfies \cite{Babaei-Aghbolagh:2022uij,Ferko:2023ruw} the $d = 4$ version of the so-called ``root-$\TT$'' flow equation,
\begin{align}\label{root_TT_flow}
    \frac{\partial \mathcal{L}^{(\gamma)}}{\partial \gamma} = \frac{1}{\sqrt{d}} \sqrt{ \widehat{T}^{(\gamma) \mu \nu} \widehat{T}_{\mu \nu}^{(\gamma)} } \, , \qquad \widehat{T}_{\mu \nu}^{(\gamma)} = T_{\mu \nu}^{(\gamma)} - \frac{1}{d} g_{\mu \nu} \tensor{T}{^{(\gamma)}^\rho_\rho} \, .
\end{align}
Here $\widehat{T}_{\mu \nu}$ is the traceless part of the stress tensor. Similar stress tensor flow equations exist for the $3d$ Born-Infeld theory \cite{Ferko:2023sps} and for a two-parameter family of $4d$ ModMax-Born-Infeld theories \cite{Ferko:2022iru,Ferko:2023ruw}; the latter family was proposed in \cite{Bandos:2020hgy}.\footnote{See also \cite{Babaei-Aghbolagh:2024uqp} for recent work on stress tensor flows for other duality-invariant theories of electrodynamics, such as the Bossard-Nicolai theory \cite{Bossard:2011ij,Carrasco:2011jv}.}

Let us remark that the two-dimensional version of the marginal root-$\TT$ flow (\ref{root_TT_flow}), which was first introduced in \cite{Ferko:2022cix} (see also \cite{Conti:2022egv,Babaei-Aghbolagh:2022leo})
is less well-understood than its irrelevant $\TT$ counterpart. For instance, it is not obvious how to unambiguously define the root-$\TT$ operator at the quantum level, although a recent proposal appeared in \cite{Hadasz:2024pew}. 
\enlargethispage{1.05\baselineskip}
Despite its uncertain quantum properties, the root-$\TT$ flow has many interesting features, besides its connection with ModMax: it preserves integrability when applied to several integrable sigma models \cite{Borsato:2022tmu,Ferko:2024ali}; admits a dimensional reduction to $1$ spacetime dimension which is related to the ``ModMax oscillator'' \cite{Garcia:2022wad,Ferko:2023ozb,Ferko:2023iha}; is related to ultra-relativistic limits, BMS, and Carrollian symmetry \cite{Rodriguez:2021tcz,Bagchi:2022nvj,Bagchi:2024unl}; and can be interpreted via modified boundary conditions in holography \cite{Ebert:2023tih,Tian:2024vln}.

Much like the status of the $2d$ root-$\TT$ deformation, it is not known how to define any stress tensor flows in $d > 2$ at the quantum level.\footnote{Likewise, it is not known whether one can fully define the ModMax theory at the quantum level, although one can study its perturbative quantization around a fixed background \cite{pinellithesis,LukeMartin:2024gsb}.} For the case of higher-dimensional $\TT$ deformations, the reason for this is that the coincident-point divergences in a point-split $\TT$ operator cannot be eliminated in a regulator-independent way, even when deforming a CFT \cite{Taylor:2018xcy}. However, this will not trouble us in the present work, since we are interested in long-distance physics that is captured by effective field theory, and we will not concern ourself with short-distance issues. In this sense, our analysis will be effectively classical. Upon restriction to this regime, there is no issue in defining classical $4d$ versions of the $\TT$ and root-$\TT$ flows, both of which are related to field-dependent metrics as we have mentioned above.

The goal of this work is to present a complementary viewpoint on the geometrization of certain $4d$ Abelian gauge theories via field-dependent metrics. Our procedure is applicable to any deformation of the $4d$ Maxwell theory by an arbitrary function of the stress tensor, or equivalently, to any $4d$ duality-invariant extension of Maxwell electrodynamics.\footnote{There exist self-dual theories which do not reduce to the Maxwell theory in any limit, such as Bialynicki-Birula electrodynamics \cite{Bialynicki-Birula:1992rcm}. We will neglect this subtlety in the present discussion.} An advantage of any geometric interpretation is that it can simplify the analysis of symmetries in the deformed theory; for instance, 
\cite{Chen:2021aid} used the field-dependent diffeomorphism associated with $2d$ $\TT$ flows to find deformed Lax connections in integrable models, which gives a characterization of the hidden symmetries in the deformed theory. Like an integrable $2d$ theory, the $4d$ Maxwell theory exhibits infinitely many conserved charges, which is clear because it is a free model. Of particular interest to us is an infinite set of conserved quantities in the Maxwell theory which descend from its $1$-form global symmetries, as explained in \cite{Hofman:2018lfz}. We will use the machinery of field-dependent metrics to study the analogue of this analysis in interacting theories of duality-invariant electrodynamics, which retain two $1$-form global symmetries. In particular, we give a geometrical (or Hodge-theoretic) characterization of conserved currents in such duality-invariant theories in terms of differential forms which are harmonic with respect to a field-dependent metric.

The structure of this proceedings contribution is as follows. In section \ref{sec:two_notions}, we introduce two ways of coupling Maxwell theory to a field-dependent metric, identify the one which we prefer, and show that the two coupling procedures agree for the ModMax theory. Section \ref{sec:duality_and_metric} then demonstrates that any duality-invariant theory can be presented as Maxwell theory coupled to a unit-determinant field-dependent metric $h_{\mu \nu}$, and vice-versa. In section \ref{sec:higher_form}, we apply these results to study the $1$-form global symmetries in theories of duality-invariant electrodynamics, and show that these symmetries can be used to construct one ordinary current $j^\mu$ for each $2$-form which is harmonic with respect to the field-dependent metric $h_{\mu \nu}$. Finally, section \ref{sec:conclusion} summarizes our results and presents directions for further research.

\section{Two Notions of Coupling to a Field-Dependent Metric}\label{sec:two_notions}

To describe what one might mean by coupling to a field-dependent metric, let us first recall the formulation of source-free Maxwell electrodynamics on a field-independent background metric. Consider the free theory of an Abelian gauge field $A_\mu$ with field strength $F_{\mu \nu} = \partial_\mu A_\nu - \partial_\nu A_\mu$ on a $4d$ spacetime manifold $\mathcal{M}$ with a metric $g_{\mu \nu}$. This theory is described by the Maxwell action in curved spacetime,
\begin{align}\label{maxwell_curved_spacetime}
    S = - \frac{1}{4} \int d^4 x \, \sqrt{g} g^{\mu \rho} g^{\nu \sigma} F_{\mu \nu} F_{\rho \sigma} \, .
\end{align}
The equation of motion which arises from (\ref{maxwell_curved_spacetime}) is
\begin{align}\label{maxwell_curved_eom}
    \partial_\mu \left( \sqrt{g} g^{\mu \rho} F_{\rho \sigma} g^{\sigma \nu} \right) = 0 \, .
\end{align}
Recall that, for any antisymmetric tensor $A^{\mu \nu}$, one has
\begin{align}\label{covariant_to_partial}
    \nabla_\mu A^{\mu \nu} = \frac{1}{\sqrt{g}} \partial_\mu \left( \sqrt{g} A^{\mu \nu} \right) \, ,
\end{align}
so the equation of motion (\ref{maxwell_curved_eom}) can be written as the covariant conservation equation
\begin{align}
    \nabla_\mu F^{\mu \nu} = 0 \, ,
\end{align}
where indices are raised or lowered with $g_{\mu \nu}$, and where $\nabla_\mu$ is the covariant derivative defined using the torsionless Levi-Civita connection.

Let us also point out that, because the Maxwell theory is invariant under conformal transformations, the curved-space action (\ref{maxwell_curved_spacetime}) on a metric $g_{\mu \nu}$ defines a theory which is equivalent to the one on a metric $\hat{g}_{\mu \nu}$ given by
\begin{align}
    \hat{g}_{\mu \nu} = e^{2 \varphi} g_{\mu \nu} \, ,
\end{align}
where $\varphi$ is a smooth function on $\mathcal{M}$. The determinant of the metric transforms as
\begin{align}
    | \det ( \tilde{g} ) | = e^{8 \varphi} | \det ( g ) |
\end{align}
under such a conformal change, so given a general metric $g_{\mu \nu}$, we may always perform a Weyl transformation with conformal factor
\begin{align}
    \varphi = - \frac{1}{8} \log \left( | \det ( g ) | \right) \, ,
\end{align}
in order to find a unit-determinant metric $\tilde{g}_{\mu \nu}$ within the same conformal class.\footnote{In order to treat both Riemannian and Lorentzian spacetimes uniformly, we will use the phrase ``unit-determiniant'' to refer to any metric $g_{\mu \nu}$ with $\det ( g ) = \pm 1$.} For this reason, we may set $g = | \det ( g ) | = 1$ without loss of generality.

Suppose we now wish to investigate the Maxwell theory on a metric that depends on the field strength $F_{\mu \nu}$, which we write as $h_{\mu \nu} = h_{\mu \nu} ( F )$. To avoid confusion, we will raise or lower indices with the background metric $g_{\mu \nu}$, and denote the inverse metric as $\left( h^{-1} \right)^{\mu \nu}$. For instance, with this convention one has
\begin{align}
    h^{\mu \nu} = g^{\mu \alpha} g^{\nu \beta} h_{\alpha \beta } \neq ( h^{-1} )^{\mu \nu} \, .
\end{align}
There are two natural, but distinct, procedures which one might follow to define ``Maxwell theory coupled to a field-dependent metric'' which are:
\begin{enumerate}[label=(\Roman*)]
    \item\label{def_one} Begin with the action (\ref{maxwell_curved_spacetime}) and replace $g_{\mu \nu}$ by $h_{\mu \nu}$, leading to the action
    \begin{align}\label{maxwell_action_field_dep_metric}
        S = - \frac{1}{4} \int d^4 x \, \sqrt{h} \left( h^{-1} \right)^{\mu \rho} \left( h^{-1} \right)^{\nu \sigma} F_{\mu \nu} F_{\rho \sigma} \, .
    \end{align}
    
    \item\label{def_two} Begin with the equation of motion (\ref{maxwell_curved_eom}) and replace $g_{\mu \nu}$ with $h_{\mu \nu}$ to write
    \begin{align}\label{maxwell_eom_field_dep_metric}
        \partial_\mu \left( \sqrt{h} \left( h^{-1} \right) ^{\mu \rho} F_{\rho \sigma} \left( h^{-1} \right) ^{\sigma \nu} \right) = 0 \, .
    \end{align}
\end{enumerate}
One might refer to procedure \ref{def_one} as ``coupling the Maxwell \emph{Lagrangian} to a field-dependent metric'' and to procedure \ref{def_two} as ``coupling the Maxwell \emph{equations of motion} to a field-dependent metric.'' The two theories defined by the procedures above are, in general, different. To see this, one can compute the Euler-Lagrange equation
\begin{align}
    \partial_\mu \left( \frac{\partial \mathcal{L}}{\partial F_{\mu \nu}} \right) = 0 \, ,
\end{align}
associated with the action (\ref{maxwell_action_field_dep_metric}), which is
\begin{align}\label{field_dep_action_eom}
    0 = \partial_\alpha &\Bigg( \sqrt{ | h | } \Big( ( h^{-1} ) ^{\rho \alpha}  ( h^{-1} ) ^{\tau \beta} F_{\rho \tau}  + \frac{\partial ( h^{-1} ) ^{\rho \sigma}}{\partial F_{\alpha \beta}} ( h^{-1} ) ^{\tau \gamma} F_{\rho \tau} F_{\sigma \gamma} \nonumber \\
    &\qquad \qquad - \frac{1}{4} h_{\mu \nu} \, \frac{\partial ( h^{-1} ) ^{\mu \nu}}{\partial F_{\alpha \beta}}  ( h^{-1} ) ^{\rho \sigma} ( h^{-1 } )^{\tau \gamma} F_{\rho \tau} F_{\sigma \gamma} \Big) \Bigg) \, .
\end{align}
The equation of motion (\ref{field_dep_action_eom}) is more complicated than (\ref{maxwell_eom_field_dep_metric}). In the remainder of this work, we will be primarily interested in studying conserved quantities, whose divergences vanish when the equations of motion are satisfied. Because we would like to have the simplest equations of motion possible, we will prefer procedure \ref{def_two} over procedure \ref{def_one}, and we will refer to the theory with equation of motion (\ref{maxwell_eom_field_dep_metric}) as ``Maxwell theory coupled to a field-dependent metric.''

We will not undertake a systematic investigation of the first procedure \ref{def_one} in this article. However, let us make a brief remark about one case where the two coupling prescriptions above actually coincide. We introduce the standard variables
\begin{align}\label{S_and_P_defn}
    S = - \frac{1}{4} F^{\mu \nu} F_{\mu \nu} \, , \qquad P = - \frac{1}{4} F^{\mu \nu} \widetilde{F}_{\mu \nu} \, .
\end{align}
Then consider the field-dependent metric $h_{\mu \nu} ( F ) $ given by
\begin{align}\label{modmax_metric}
    h_{\mu \nu} ( F ) = \cosh ( \gamma ) g_{\mu \nu} + \frac{\sinh ( \gamma )}{\sqrt{S^2 + P^2}} \left(  S g_{\mu \nu} - \tensor{F}{_\mu^\rho} \tensor{F}{_\rho_\nu} \right) \, .
\end{align}
Assuming $| \det ( g ) | = 1$, one finds by direct calculation that
\begin{align}
    | \det ( h_{\mu \nu} ) | = 1 \, ,
\end{align}
and that the inverse metric is
\begin{align}\label{modmax_inverse_metric}
    \left( h^{-1} \right)^{\mu \nu} = \cosh ( \gamma ) g^{\mu \nu} -  \frac{\sinh ( \gamma )}{\sqrt{S^2 + P^2}} \left(  S g^{\mu \nu} - F^{\mu \rho} \tensor{F}{_\rho^\nu}  \right) \, .
\end{align}
Furthermore, this metric has the property that
\begin{align}\label{condition_lag_eom_same}
    \frac{\partial ( h^{-1} )^{\rho \sigma}}{\partial F_{\alpha \beta}} ( h^{-1} )^{\tau \gamma} F_{\rho \tau} F_{\sigma \gamma} = 0 \, .
\end{align}
This means that the equation of motion (\ref{field_dep_action_eom}) arising from procedure \ref{def_one}, for this special choice of metric, reduces to 
\begin{align}\label{field_dep_action_eom_simplified}
    0 = \partial_\alpha \left(  ( h^{-1} )^{\alpha \rho}   F_{\rho \tau} ( h^{-1} )^{\tau \beta} \right) \, ,
\end{align}
which is the same as the equation of motion (\ref{maxwell_eom_field_dep_metric}) arising from procedure \ref{def_two}.

Since the two procedures agree in this case, we can identify the Lagrangian for the Maxwell theory coupled to the field-dependent metric (\ref{modmax_metric}) by computing
\begin{align}
     - \frac{1}{4} ( h^{-1} ) ^{\mu \rho}  ( h^{-1} )^{\nu \sigma}  F_{\mu \nu} F_{\rho \sigma} = \cosh \left( 2 \gamma \right) S + \sinh ( 2 \gamma ) \sqrt{ S^2 + P^2 } \, ,
\end{align}
and thus the Maxwell theory on this field-dependent metric corresponds to the ModMax theory (\ref{modmax}) at parameter $2 \gamma$,
\begin{align}
    S_{\text{ModMax,} 2 \gamma} = \int d^4 x \, \sqrt{ g } \, \left( \cosh \left( 2 \gamma \right) S + \sinh ( 2 \gamma ) \sqrt{ S^2 + P^2 } \right) \, .
\end{align}
Therefore, one can generate the ModMax theory by coupling the Maxwell theory to a field-dependent metric, and in this case it does not matter whether one performs this coupling in the action or in the equations of motion. This metric coupling is reminiscent of the ``deformation map'' introduced in \cite{Garcia:2022wad}, which likewise transforms the isotropic $2d$ harmonic oscillator into the ModMax oscillator at parameter $2 \gamma$.

\section{Duality Invariance and Field-Dependent Metrics}\label{sec:duality_and_metric}

We have now seen that ModMax is one example of a model for which the equations of motion can be represented via a coupling to a field-dependent metric. Perhaps this does not come as a surprise, since it is known that certain deformations can be represented by coupling a simpler theory to a gravitational sector. For instance, any family of $4d$ theories $\mathcal{L}^{(\lambda)} ( S, P )$ obtained from a classical $\TT$-like deformation of a seed theory $\mathcal{L}^{(0)} ( S, P ) $ can be understood via a metric approach, as investigated in \cite{Conti:2022egv,Morone:2024ffm}. Recent work suggests that this is also possible for root-$\TT$ deformations \cite{Babaei-Aghbolagh:2024hti,Tsolakidis:2024wut}. It is natural to ask which other theories admit such a description.

The question we will now address is: for which theories of source-free non-linear electrodynamics $\mathcal{L} ( S, P )$ are the equations of motion equivalent to 
\begin{align}\label{question_fdm}
    \partial_\mu \left( \left( h^{-1} \right)^{\mu \rho} F_{\rho \sigma} \left( h^{-1} \right)^{\sigma \nu} \right) = 0 \, ,
\end{align}
for a field-dependent metric $h_{\mu \nu} ( F )$ with constant determinant, $| \det ( h ) | = c$?

In this question, the form (\ref{question_fdm}) of the proposed equation of motion is the same as (\ref{maxwell_eom_field_dep_metric}) in the special case where the determinant of the field-dependent metric is constant.\footnote{If $| \det ( h ) | = c$, we can perform a rescaling $h_{\mu \nu} \to \lambda h_{\mu \nu}$, which does not affect (\ref{question_fdm}), to take $c = 1$ without loss of generality. However it will be convenient to leave the constant $c$ arbitrary for now.} We impose this condition for simplicity. For instance, in view of equation (\ref{covariant_to_partial}), when $| \det ( h ) |$ is constant we have $\nabla_\mu A^{\mu \nu} = \partial_\mu A^{\mu \nu}$, and we need not distinguish between ordinary and covariant divergences of antisymmetric tensors.

To look for other examples for which the answer to this question is affirmative (besides ModMax), we now parameterize the class of theories we will consider, and a corresponding family of field-dependent metrics. We reiterate that we consider only source-free theories of $4d$ electrodynamics described by Lagrangians which are Lorentz scalars constructed from $F_{\mu \nu}$ but not from its derivatives. There are two useful ways to parameterize such Lagrangians. One is using the two variables $S$ and $P$ of equation (\ref{S_and_P_defn}). The other parameterization is in terms of the two variables
\begin{align}
    x_1 = \tr ( F^2 ) = \tensor{F}{^\mu_\rho} \tensor{F}{^\rho_\mu} \, , \qquad x_2 = \tr ( F^4 ) = \tensor{F}{^\mu_\rho} \tensor{F}{^\rho_\sigma} \tensor{F}{^\sigma_\tau} \tensor{F}{^\tau_\mu} \, .
\end{align}
Either pair of variables, $(S, P)$ or $(x_1, x_2)$, provides a functionally complete set of Lorentz invariants constructed from $F_{\mu \nu}$ in $4d$, and the two sets are related by
\begin{align}\label{x_to_S_P}
    x_1 = 4 S \, , \qquad x_2 = 4 P^2 + 8 S^2 \, .
\end{align}
Let us pause to comment on the spacetime signature. We will alternately work on either a Riemannian spacetime manifold, with Euclidean signature $(+, +, +, +)$, or on a Lorentzian spacetime with signature $(-, +, +, +)$. One or the other of these choices may be more convenient, depending on which aspects of a given theory one would like to discuss. To understand a model of electrodynamics as a physically reasonable field theory, for instance describing the propagation of (possibly self-interacting) wavelike disturbances in time, one should adopt Lorentzian signature. The choice of Lorentz signature also appears to be more natural for the purpose of constructing certain consistent and manifestly duality-invariant presentations of gauge theories in $4d$ or chiral form theories in other dimensions, such as the $6d$  Pasti-Sorokin-Tonin (PST) formulation \cite{Pasti:1995tn,Pasti:1996vs,Pasti:1997gx} (or its generalization to higher dimensions \cite{Buratti:2019guq}), the $4d$ Ivanov-Zupnik formalism \cite{Ivanov:2002ab,Ivanov:2003uj} (and its higher-$d$ extension \cite{Kuzenko:2019nlm}), and the related Ivanov-Nurmagambetov-Zupnik \cite{Ivanov:2014nya} representation. On the other hand, the choice of Euclidean signature can be more useful for emphasizing topological field theory aspects of a model. For this reason we will use Euclidean signature in section \ref{sec:higher_form}, since the discussion of $1$-form symmetries is closely connected to the existence of topologically protected line operators in the theory.

In this section, to write formulas which are valid for either choice of spacetime signature, we will use a definition of the Hodge star which involves a factor of $\sqrt{ - \det ( g ) }$. This introduces a factor of $i$ in Euclidean signature; for example, if $F$ is a real field strength, then $\ast F$ is purely imaginary on a Riemannian manifold. The advantage of this definitional choice is that the formula (\ref{x_to_S_P}) holds in either spacetime signature, so we may treat both cases at once.

Having adopted this convention, we can interchangeably parameterize a general Lagrangian for a theory of non-linear electrodynamics as
\begin{align}
    \mathcal{L} = \mathcal{L} ( S,  P ) = \mathcal{L} ( x_1, x_2 ) \, .
\end{align}
It is convenient to record the condition for electric-magnetic duality invariance in both sets of variables. We define symbols $\mathcal{D} ( S, P )$ and $\mathcal{D} ( x_1, x_2 )$ so that the self-duality equation reads $\mathcal{D} = 1$ in either parameterization. In terms of $S$ and $P$, a Lagrangian $\mathcal{L} ( S, P )$ describes a duality-invariant theory if
\begin{align}\label{duality_invariance_S_P_variables}
    \mathcal{D} ( S, P ) = \mathcal{L}_S^2 - \frac{2 S}{P} \mathcal{L}_S \mathcal{L}_P - \mathcal{L}_P^2 = 1 \, , 
\end{align}
whereas in $(x_1, x_2)$ variables, the condition is
\begin{align}\label{duality_invariance_x_variables}
    \mathcal{D} ( x_1, x_2 ) = 16 \mathcal{L}_1^2 + 16 x_1 \mathcal{L}_1 \mathcal{L}_2 - 8 \left( 2 x_2 - x_1^2 \right) \mathcal{L}_2^2 = 1 \, .
\end{align}
Here subscripts indicate partial derivatives with respect to the corresponding argument, and we have defined the shorthand
\begin{align}
    \mathcal{L}_1 = \frac{\partial \mathcal{L}}{\partial x_1} \, , \qquad \mathcal{L}_2 = \frac{\partial \mathcal{L}}{\partial x_2} \, .
\end{align}
To describe the general class of field-dependent metrics which we will allow, it will be more convenient to first specify the inverse metric $\left( h^{-1} \right)^{\mu \nu}$, and then allow the metric itself to be defined implicitly in terms of its inverse. Consider the ansatz
\begin{align}\label{inverse_metric_defn}
    \left( h^{-1} \right)^{\mu \nu} = A ( x_1, x_2 ) g^{\mu \nu} + B ( x_1, x_2 ) \tensor{F}{^\mu_\rho} \tensor{F}{^\rho^\nu} \, .
\end{align}
As above, all indices will be raised or lowered with $g_{\mu \nu}$. The following calculation will not depend on the spacetime signature or the precise form of $g_{\mu \nu}$, so long as $| \det ( g ) | = 1$. Recall that $g_{\mu \nu}$ is the fixed, field-independent background metric which defines the undeformed Maxwell theory (\ref{maxwell_curved_spacetime}). We have argued that we may take $| \det ( g ) | = 1$ without loss of generality due to conformal invariance, so we assume this in what follows. However, the equations of motion (\ref{question_fdm}) with metric (\ref{inverse_metric_defn}) are generally \emph{not} invariant under conformal transformations because these transformations act non-trivially on the arguments $x_1$, $x_2$ of the functions $A$ and $B$.

One might ask why we do not allow for additional terms in (\ref{inverse_metric_defn}), such as
\begin{align}\label{additional_term}
    C ( x_1, x_2 ) \tensor{F}{^\mu_\rho} \tensor{F}{^\rho_\sigma} \tensor{F}{^\sigma_\tau} \tensor{F}{^\tau^\nu} \, .
\end{align}
The reason is that all factors of the inverse metric will eventually be contracted with field strength tensors. If we had allowed an additional term (\ref{additional_term}) in the metric, then upon contraction with $\tensor{F}{_\nu^\alpha}$, we would generate a contribution
\begin{align}
    C ( x_1, x_2 ) \tensor{F}{^\mu_\rho} \tensor{F}{^\rho_\sigma} \tensor{F}{^\sigma_\tau} \tensor{F}{^\tau^\nu} \tensor{F}{_\nu^\alpha} = C ( x_1, x_2 ) \big( F^5 \big) ^{\mu \alpha} \, ,
\end{align}
where we use the notation
\begin{align}
    \tensor{\left( F^n \right)}{^\mu_\nu} = \underbrace{\tensor{F}{^\mu_{\alpha_1}} \tensor{F}{^{\alpha_1}_{\alpha_2}} \ldots \tensor{F}{^{\alpha_{n-2}}_{\alpha_{n-1}}} \tensor{F}{^{\alpha_{n-1}}_\nu}}_{n \text{ times}} \, ,
\end{align}
for the matrix product of $n$ factors of $F$.

However, the Cayley-Hamilton determines the fifth power of the matrix $F$ as
\begin{align}\label{CH_F5}
    0 = \tensor{\big( F^5 \big)}{^\mu_\nu} - \frac{1}{2} x_1 \tensor{\left( F^3 \right)}{^\mu_\nu} + \frac{1}{4} \left( \frac{1}{2} x_1^2 - x_2 \right) \tensor{F}{^\mu_\nu} \, .
\end{align}
Therefore, the effect of adding a term (\ref{additional_term}) to the metric -- after contraction with a factor of the field strength -- can be absorbed into the functions $A$ and $B$ multiplying the lower-order terms in (\ref{inverse_metric_defn}). Thus (\ref{inverse_metric_defn}) is effectively the most general symmetric tensor constructed only from $g^{\mu \nu}$ and $F^{\mu \nu}$ which one could allow.

Recall that our question stipulates that the determinant of $h_{\mu \nu}$ should be constant. Because $\det ( h^{-1} ) \det ( h ) = 1$, in order to investigate the condition that $\det ( h )$ is constant, it suffices to compute $\det ( h^{-1} )$. By an explicit calculation one finds
\begin{align}\label{det_h_ansatz}
    | \det ( h^{-1} ) | = \frac{1}{64} \left( 8 A^2 + 4 A B x_1 + B^2 \left( x_1^2 - 2 x_2 \right) \right)^2 \, , 
\end{align}
and thus $\left| \det ( h ) \right| = c$, for a positive constant $c$, if and only if
\begin{align}
    \left| A^2 + \frac{1}{2} A B x_1 + \frac{1}{8} B^2 \left( x_1^2 - 2 x_2 \right) \right| = c^{-1/2} \, .
\end{align}
Let us now turn to the equations of motion. We begin with a theory of electrodynamics $\mathcal{L} ( x_1, x_2 )$ on a field-independent metric. The Euler-Lagrange equation is
\begin{align}
    0 &= \partial_\mu \left( \frac{\partial \mathcal{L}}{\partial F_{\mu \nu}} \right) \, ,
\end{align}
which for $| \det ( g ) | = 1$, can be written as
\begin{align}\label{nled_eom}
    0 &= \partial_\mu \left( \mathcal{L}_1 F^{\mu \nu} + 2 \mathcal{L}_2 \left( F^3 \right)^{\mu \nu} \right) \, . 
\end{align}
We compare this to the equation of motion for the Maxwell theory coupled to a unit-determinant field-dependent metric, which is
\begin{align}\label{maxwell_metric_eom}
    \partial_\mu \left( ( h^{-1} ) ^{\mu \rho} F_{\rho \sigma} ( h^{-1} )^{\sigma \nu} \right) = 0 \, .
\end{align}
We would like the equations of motion (\ref{maxwell_metric_eom}) to agree with (\ref{nled_eom}). To do this, we can impose that the quantities acted upon by the derivative $\partial_\mu$ in these two expressions are proportional to one another. That is, we require
\begin{align}
    ( h^{-1} )^{\mu \rho} F_{\rho \sigma} (h^{-1})^{\sigma \nu} = a \left( F^{\mu \nu} + 2 \left( F^3 \right)^{\mu \nu} \right) \, ,
\end{align}
where $a$ is constant. Expanding the left side using the definition of $( h^{-1} )^{\mu \nu}$, and simplifying with the Cayley-Hamilton theorem in the form (\ref{CH_F5}), this becomes
\begin{align}\hspace{-5pt}
    \left( A^2 + \frac{1}{8} B^2 ( 2 x_2 - x_1^2 ) \right) F^{\mu \nu} + \left( 2 A B + \frac{1}{2} B^2 x_1 \right) \left( F^3 \right)^{\mu \nu} = a \left( \mathcal{L}_1 F^{\mu \nu} + 2 \mathcal{L}_2 \left( F^3 \right)^{\mu \nu} \right) \, .
\end{align}
Demanding the coefficients of independent powers of $F$ to agree gives
\begin{align}\label{coeffs_agree}
    a \mathcal{L}_1 = A^2 + \frac{1}{8} B^2 ( 2 x_2 - x_1^2 )  \, , \qquad a \mathcal{L}_2 &=  A B + \frac{1}{4} B^2 x_1 \, .
\end{align}
We note that (\ref{coeffs_agree}) is a system of two algebraic equations for the two unknown functions $A$, $B$, and thus it generically admits a solution. However, we are not guaranteed that this solution will respect the constant-determinant condition. It turns out that this constant-determinant condition is related to duality invariance. Indeed, substituting the solutions (\ref{coeffs_agree}) for the derivatives of the Lagrangian into the left side of the condition (\ref{duality_invariance_x_variables}) for duality invariance, one finds
\begin{align}
    16 \mathcal{L}_1^2 + 16 x_1 \mathcal{L}_1 \mathcal{L}_2 - 8 \left( 2 x_2 - x_1^2 \right) \mathcal{L}_2^2 &= \frac{1}{4 a^2} \left( 8 A^2 + 4 A B x_1 + B^2 \left( x_1^2 - 2 x_2 \right) \right)^2 \, ,
\end{align}
and comparing to the expression (\ref{det_h_ansatz}) for the determinant of the inverse metric gives
\begin{align}\label{duality_equals_determinant}
    \mathcal{D} ( x_1, x_2 ) &= \frac{16}{a^2} | \det \left( h^{-1} \right) | \, .
\end{align}
If $\mathcal{L}$ enjoys duality-invariance, then the left side of (\ref{duality_equals_determinant}) equals $1$, so $| \det \left( h^{-1} \right) | = \frac{a^2}{16}$. The constant $a$ was an arbitrary proportionality factor between two equations of motion, which we can choose to be $a = 4$ without loss of generality, so in this case duality invariance and equivalence of the equations of motion implies $| \det ( h^{-1} ) | = 1$ and thus $| \det ( h ) | = 1$. Conversely, if $| \det \left( h^{-1} \right) | = c^{-1}$, we may choose to take $a = 4 \sqrt{c^{-1}}$, which implies that $\mathcal{L}$ satisfies the condition (\ref{duality_invariance_x_variables}) for duality invariance. Again, by performing a rescaling of the metric $h_{\mu \nu}$, we can choose to take $c = 1$.

We conclude that the equations of motion for a theory of non-linear electrodynamics are equivalent to those for the Maxwell theory coupled to a field-dependent background metric $h_{\mu \nu}$ with $| \det ( h ) | = 1$ if and only if the theory is electric-magnetic duality-invariant.

\subsubsection*{\ul{\it Relation to stress tensor deformations}}

Theories of duality-invariant non-linear electrodynamics can be alternatively characterized in terms of deformations of the Lagrangian which are constructed from the energy-momentum tensor, as described in \cite{Ferko:2023wyi}. We will now briefly review the results of that work, to which we refer the reader for further details.

Let $\mathcal{L}^{(\lambda)} ( S , P )$ -- or equivalently $\mathcal{L}^{(\lambda)} ( x_1, x_2 )$ -- be a one-parameter family of Lagrangians which obey the duality-invariance condition (\ref{duality_invariance_S_P_variables}) or (\ref{duality_invariance_x_variables}). Then it was shown in \cite{Ferko:2023wyi} that this family of Lagrangians satisfies a differential equation
\begin{align}\label{stress_tensor_flow}
    \frac{\partial \mathcal{L}^{(\lambda)}}{\partial \lambda} = f \left( \lambda, T_{\mu \nu}^{(\lambda)} \right) \, ,
\end{align}
where $f$ is a Lorentz scalar function constructed from the energy-momentum tensor
\begin{align}
    T_{\mu \nu}^{(\lambda)} = - \frac{2}{\sqrt{g}} \frac{\delta S^{( \lambda )}  }{\delta g^{\mu \nu}} \, , \quad S^{(\lambda)} = \int d^4 x \, \mathcal{L}^{(\lambda)} \, , 
\end{align}
of the theory. The function $f$ may also have explicit dependence on $\lambda$. It was also shown that, conversely, given any such function $f \left( \lambda, T_{\mu \nu}^{(\lambda)} \right) $ and a duality-invariant initial condition $\mathcal{L}^{( \lambda = 0 )}$, such as the Maxwell theory, the differential equation (\ref{stress_tensor_flow}) defines a one-parameter family of duality-invariant Lagrangians.\footnote{Analogous statements have also been shown for chiral tensor theories in six spacetime dimensions \cite{Ferko:2024zth} and for Lorentz-invariant theories of chiral bosons in two dimensions \cite{Ebert:2024zwv}.}

On the other hand, we have seen that the equations of motion for any duality-invariant model can be understood as those of the curved-space Maxwell theory, with the metric $g_{\mu \nu}$ replaced by a unit-determinant field-dependent metric $h_{\mu \nu}$.

Combining these results, we find that the following conditions are equivalent:

\begin{enumerate}[label=(\roman*)]
    \item $\mathcal{L}^{(\lambda)} ( S, P )$ is a family of theories of duality-invariant electrodynamics;

    \item $\mathcal{L}^{(\lambda)} ( S, P )$ satisfies a differential equation in the parameter $\lambda$ which is driven by a function of the energy-momentum tensor, with a duality-invariant initial condition $\mathcal{L}^{(\lambda_0)} ( S, P )$ for some $\lambda_0$; and

    \item the equations of motion associated with $\mathcal{L}^{(\lambda)} (  S, P )$ take the form (\ref{maxwell_eom_field_dep_metric}) for a family of field-dependent metrics $h_{\mu \nu}^{(\lambda)} ( F )$ with $\left| \det ( h^{(\lambda)} ( F ) ) \right| = 1$.
\end{enumerate}
We conclude that a general stress tensor deformation of the Maxwell theory in four dimensions can be geometrized in terms of a field-dependent metric. In a sense, this extends the results of \cite{Conti:2022egv,Morone:2024ffm,Babaei-Aghbolagh:2024hti,Tsolakidis:2024wut}, which give geometrical realizations of classical deformations by the four-dimensional $\TT$ and root-$\TT$ operators, to the case of deformations by arbitrary functions of $T_{\mu \nu}$.

However, we should point out that the geometrization approach which we have taken here differs from the strategy of works such as \cite{Morone:2024ffm}. In particular, our procedure couples the equations of motion for a seed theory (i.e. Maxwell) to an explicit, fixed field-dependent metric $h_{\mu \nu}$, whereas \cite{Morone:2024ffm} couples the seed theory to dynamical gravity in the Palatini formalism, with a particular action for the gravity sector.

\section{Application to Higher-Form Symmetries}\label{sec:higher_form}

We now turn to the study of $1$-form global symmetries in theories of duality-invariant electrodynamics, and in particular to the construction of ordinary conserved currents from these higher-form symmetries. This offers an application of the field-dependent metric machinery which was developed in sections \ref{sec:two_notions} and \ref{sec:duality_and_metric}. First, let us review some of the basic definitions and facts about higher-form symmetries.

We say that a theory has a $p$-form global symmetry if there exists a totally antisymmetric $(p+1)$-form current $j^{\mu_1 \ldots \mu_{p+1}}$ which satisfies the conservation equation
\begin{align}
    \partial_{\mu_1} j^{\mu_1 \ldots \mu_{p+1}} = 0 \, .
\end{align}
In this language, an ordinary symmetry with a conserved one-form Noether current $j^\mu$ is referred to as a $0$-form global symmetry. The Maxwell theory with Lagrangian $\mathcal{L} = - \frac{1}{4} F_{\mu \nu} F^{\mu \nu}$ enjoys two separate one-form global symmetries, which we refer to as the ``electric'' and ``magnetic'' one-form global symmetries. The conserved $2$-form current associated with the electric $1$-form global symmetry is
\begin{align}\label{electric_one_form}
    j_{\mu \nu}^{(E)} = F_{\mu \nu} \, ,
\end{align}
which is conserved by virtue of the source-free equation of motion $\partial_\mu F^{\mu \nu} = 0$, and the $2$-form current associated with the magnetic $1$-form global symmetry is
\begin{align}\label{magnetic_one_form}
    j_{\mu \nu}^{(M)} = \widetilde{F}_{\mu \nu} \, ,
\end{align}
which is conserved due to the Bianchi identity $\partial_\mu \widetilde{F}^{\mu \nu} = 0$.

The presence of a $1$-form global symmetry signals the existence of a topologically protected line operator (also referred to as a symmetry defect operator) in a quantum field theory. For instance, in the Maxwell conformal field theory, the objects which are charged under the electric $1$-form symmetry are the Wilson lines, and the line operators charged under the magnetic $1$-form symmetry are the 't Hooft lines. In vacuum, the symmetry group for both the electric and magnetic $1$-form symmetries is $U(1)$, but these are broken to a discrete $\mathbb{Z}_N$ subgroup by the inclusion of quantized electric charges and dynamical magnetic monopoles, respectively.

Unlike the Maxwell theory, which is a microscopically complete CFT, we will study $1$-form global symmetries in general theories of duality-invariant electrodynamics, which are merely effective field theories. An example to keep in mind is the Born-Infeld model, with Lagrangian (\ref{born_infeld}). The Born-Infeld equations of motion can be derived by studying open string propagation in a D-brane background and demanding the vanishing of the one-loop worldsheet beta function, in the approximation where the derivatives $\partial_\rho F_{\mu \nu}$ of the field strength are small \cite{Abouelsaood:1986gd}. Therefore, in the example of Born-Infeld, we are interested in studying this theory in a regime where the approximation of ignoring derivatives of field strengths is valid. However, we should note that when the length scale set by typical field strengths $F_{\mu \nu}$ is small compared to the length scale set by the tension $T$ in equation (\ref{born_infeld}), the Born-Infeld theory reduces to Maxwell electrodynamics. So to be precise, for the Born-Infeld example, we are interested in studying the theory in a regime where \emph{derivatives} of the field strength are taken small, but the \emph{magnitude} of the field strength is not so small that the dynamics are well-described by Maxwell electrodynamics. More generally, for other theories $\mathcal{L} ( S, P )$ of duality-invariant electrodynamics besides Born-Infeld, we assume that there exists some regime in which $\mathcal{L}$ is a good low-energy effective field theory, and we restrict attention to this regime.\footnote{The effective field theory origin of ModMax is not fully understood, but it may arise from a coupling to an axio-dilaton auxiliary scalar field \cite{Lechner:2022qhb}, via a brane-like construction similar to Dirac-Born-Infeld \cite{Nastase:2021ffq}, or from a class of models with specific higher-derivative interactions \cite{Kuzenko:2024zra}.}

We are not the first to consider generalized global symmetries in theories of duality-invariant electrodynamics. For instance, the $1$-form symmetries of ModMax and Born-Infeld theory were considered in \cite{Chatzistavrakidis:2021dqg}, where the 't Hooft anomalies associated with these symmetries were also studied. See also \cite{Cordova:2023ent} for a discussion of the relationship between electric-magnetic duality invariance and generalized global symmetries in the Maxwell theory. The novelty in the present analysis is in the application of field-dependent metrics to the study of these symmetries, and in particular, to analyze the construction of lower-form currents from higher-form currents, which was performed in \cite{Hofman:2018lfz} for the Maxwell theory.

\subsection{One-Form Symmetries in Non-Linear Electrodynamics}

As we have reviewed, the electric and magnetic $1$-form global symmetries of the Maxwell theory (\ref{electric_one_form}) and (\ref{magnetic_one_form}) are a consequence of the source-free equations of motion and the Bianchi identity. In this case, the special properties of the Maxwell theory -- namely, that it is a CFT -- imply that these two higher-form currents are Hodge dual to one another. More generally, in any $4d$ conformal field theory, it is believed that the Hodge dual of any conserved $2$-form current must also be conserved, which is a phenomenon known as photonization \cite{Hofman:2018lfz}.

In more general theories of non-linear electrodynamics, viewed as effective field theories, there still exists a pair of conserved two-form currents, although they will not be Hodge dual to one another. Given any Lagrangian $\mathcal{L} ( S, P )$ in the class which we are considering, one can define
\begin{align}\label{Gt_def}
    \widetilde{G}^{\mu \nu} = 2 \frac{\partial \mathcal{L}}{\partial F_{\mu \nu}} \, .
\end{align}
Any such theory of electrodynamics enjoys two $1$-form symmetries with currents
\begin{align}
    j_{\mu \nu}^{(E)} = \widetilde{G}_{\mu \nu} \, , \qquad j_{\mu \nu}^{(M)} = \widetilde{F}_{\mu \nu} \, ,
\end{align}
which are again conserved by virtue of the equations of motion and the Bianchi identity, respectively. In terms of (\ref{Gt_def}), the condition (\ref{duality_invariance_S_P_variables}) for duality invariance reads
\begin{align}\label{BB_condition}
    \widetilde{G}^{\mu \nu} G_{\mu \nu} + \widetilde{F}^{\mu \nu} F_{\mu \nu} = 0 \, ,
\end{align}
which is the form of the duality invariance condition studied in \cite{BialynickiBirula:1984tx}, and is equivalent to the condition stated by Gaillard and Zumino \cite{Gaillard:1981rj}.

Let us now discuss how lower-form symmetries can be constructed from higher-form symmetries. Given any theory which exhibits a $p$-form global symmetry, associated with a $(p+1)$-form conserved current $j^{\mu_1 \ldots \mu_{p+1}}$, one can construct a $p$-form conserved current
\begin{align}\label{derived_conserved_current}
    j^{\mu_1 \ldots \mu_p} = j^{\mu_1 \ldots \mu_{p+1}} \partial_{\mu_{p+1}} f
\end{align}
for any scalar function $f$. The resulting $p$-form current $j^{\mu_1 \ldots \mu_p}$ is trivially conserved due to the conservation and antisymmetry of $j^{\mu_1 \ldots \mu_{p+1}}$. We say that this derived $p$-form current is trivial because the conservation of this current does not imply any new symmetries, beyond the one which was already present and led to the conservation of the ``parent'' current $j^{\mu_1 \ldots \mu_{p+1}}$.

For instance, in any theory of non-linear electrodynamics, one can construct the conserved $0$-form currents
\begin{align}\label{trivial_currents}
    j^\mu = \widetilde{G}^{\mu \nu} \partial_\nu g + \widetilde{F}^{\mu \nu} \partial_\nu f \, ,
\end{align}
for any scalar functions $g$ and $f$. When $f = 0$, one can show that the current (\ref{trivial_currents}) is the conserved Noether current associated with gauge transformations $A \to A + d \lambda$, whose charge acts trivially on gauge-invariant quantities and is therefore not physically interesting. When $g = 0$, the current (\ref{trivial_currents}) can be interpreted as the Noether current associated with gauge transformations of the ``dual photon'' field $B_\mu$. That is, since $\partial_\mu \widetilde{G}^{\mu \nu} = 0$, which implies that $d G = 0$, the $2$-form $G$ can be trivialized as
\begin{align}
    G = d B
\end{align}
for a one-form $B$, and then dual gauge transformations of the form $B \to B + d \lambda$ are symmetries of the theory, albeit also trivial ones.

By taking specific choices of the functions $f$ and $g$, one can find examples of the currents (\ref{trivial_currents}) whose triviality is not immediately obvious. For instance, in the ModMax theory, one can define the angular variables
\begin{align}
    \cos ( \varphi ) = \frac{S}{\sqrt{S^2 + P^2}} \, , \qquad 
    \sin ( \varphi ) = \frac{P}{\sqrt{S^2 + P^2}}  \, ,
\end{align}
and then note that for any function $\hat{f}$, the current
\begin{align}\label{modmax_current_two}
    j^\mu = F^{\mu \nu} \partial_\nu \hat{f} ( \varphi ) \, ,
\end{align}
is conserved. However, the currents (\ref{modmax_current_two}) arise from a particular choice of the currents (\ref{trivial_currents}) where $\hat{f}$ is related to $f$ and $g$, so these currents are associated with trivial gauge symmetries and thus their charges do not act on gauge-invariant quantities.

\subsection{Construction of Non-Trivial Currents in Maxwell Theory}

One might ask if there is some way to modify the construction of the currents (\ref{derived_conserved_current}) so that they are no longer trivial. Indeed this is possible in some cases. The triviality of the currents (\ref{derived_conserved_current}) is a consequence of the fact that the higher-form current is contracted with the gradient of a scalar, which is an exact one-form. Suppose that one instead contracts the higher-form current against a one-form which is not exact,
\begin{align}\label{try_non_trivial_current}
    j^{\mu_1 \ldots \mu_p} = j^{\mu_1 \ldots \mu_{p+1}}  \Lambda_{\mu_{p+1}} \, ,
\end{align}
where $\Lambda_{\mu_{p+1}} \neq \partial_{\mu_{p+1}} f$. The resulting $p$-form current is conserved if
\begin{align}
    \partial_{\mu_{p}} j^{\mu_1 \ldots \mu_p} = j^{\mu_1 \ldots \mu_{p+1}} \partial_{\mu_{p}} \Lambda_{\mu_{p+1}} = 0 \, .
\end{align}
This is a differential constraint which relates the components of the $p$-form current to the components of $\Lambda$. Such a constraint is difficult to solve in general, and typically requires the components of $\Lambda$ to depend on the fields of the theory.

However, in certain situations, this procedure simplifies. Suppose that we return to the Maxwell theory with conserved two-form currents $j_{\mu \nu}^{(E)} = F_{\mu \nu}$ and $j_{\mu \nu}^{(M)} = \widetilde{F}_{\mu \nu}$. One can make an ansatz for a $0$-form current which takes the form
\begin{align}\label{maxwell_ansatz}
    j^\mu = F^{\mu \nu} \Lambda_\nu^{(E)} + \widetilde{F}^{\mu \nu} \Lambda_\nu^{(M)} \, , 
\end{align}
where neither $\Lambda_\mu^{(E)}$ nor $\Lambda_{\mu}^{(M)}$ is exact. Let us investigate the condition for conservation of this current. One finds that
\begin{align}
    \partial_\mu j^\mu = F^{\mu \nu} \partial_\mu \Lambda_\nu^{(E)} + \widetilde{F}^{\mu \nu} \partial_\mu \Lambda_\nu^{(M)} \, ,
\end{align}
where we have used the equations of motion and Bianchi identity, which vanishes if
\begin{align}\label{maxwell_vanish_intermediate}
    F \wedge \ast d \Lambda^{(E)} + \widetilde{F} \wedge \ast d \Lambda^{(M)} = 0 \, .
\end{align}
In this section, we work in Euclidean\footnote{The analysis of this section can be converted to Lorentzian signature, where $\ast \ast = -1$ when acting on $2$-forms, by including factors of $i$ in appropriate places in all of the formulas.} signature where the definition of the Hodge star involves a factor of $\sqrt{ | \det ( g ) | }$ rather than $\sqrt{ - \det ( g ) }$, so that $\ast \ast = 1$ when acting on $2$-forms. With these conventions, (\ref{maxwell_vanish_intermediate}) is equivalent to
\begin{align}\label{current_cons_lambdas_condition}
    F \wedge \left( d \Lambda^{(M)} + \ast d \Lambda^{(E)} \right) = 0 \, ,
\end{align}
which can be solved by taking
\begin{align}\label{harmonic_solution}
    d \Lambda^{(M)} = - \ast d \Lambda^{(E)} \, .
\end{align}
The relation (\ref{harmonic_solution}) is sometimes called a twisted self-duality condition.\footnote{See \cite{Avetisyan:2022zza} for a democratic formulation of Lagrangians for chiral fields which incorporates such twisted self-duality constraints on the potentials.} If we choose a solution of the type (\ref{harmonic_solution}), then both $d \Lambda^{(E)}$ and $d \Lambda^{(M)}$ are harmonic:
\begin{align}\label{harmonic_E_and_M}
    0 = d \left( d \Lambda^{(E)} \right) = d \left( \ast d \Lambda^{(E)} \right) \, , \qquad 0 = d \left( d \Lambda^{(M)} \right) = d \left( \ast d \Lambda^{(M)} \right) \, .
\end{align}
One can decompose $d \Lambda^{(E)}$, $d \Lambda^{(M)}$ into self-dual and anti-self-dual parts $\big( d \Lambda^{(E)} \big)^{\pm}$, $\big( d \Lambda^{(M)} \big)^{\pm}$, where for any two-form $A$ in Euclidean signature, we define
\begin{align}
    A^{\pm} = \frac{1}{2} \left( 1 \pm \ast \right) A \, .
\end{align}
Then the harmonicity condition (\ref{harmonic_E_and_M}) implies that each of the self-dual and anti-self-dual parts are separately closed,
\begin{align}
    d \left( d \Lambda^{(E)} \right)^+ = d \left( d \Lambda^{(E)} \right)^- = 0 \, , \quad d \left( d \Lambda^{(M)} \right)^+ = d \left( d \Lambda^{(M)} \right)^- = 0 \, .
\end{align}
Hence one can solve equation (\ref{harmonic_solution}) by taking
\begin{align}
    d \Lambda^{(E)} = \left( d \Lambda^{(M)} \right)^- - \left( d \Lambda^{(M)} \right)^+ \, .
\end{align}
The upshot of this calculation is that we may now identify two independent families of solutions which lead to non-trivial conserved charges in Maxwell theory, following \cite{Hofman:2018lfz}. First, given any non-exact one-form $\Lambda$ with self-dual derivative,
\begin{align}
    \ast d \Lambda = d \Lambda \, ,
\end{align}
we can choose $\Lambda^{(M)} = \Lambda$ and $\Lambda^{(E)} = - \Lambda$, and this gives a solution to (\ref{maxwell_ansatz}). Conversely, given any non-exact one-form $\overbar{\Lambda}$ with anti-self-dual derivative,
\begin{align}
    \ast d \overbar{\Lambda} = - d \overbar{\Lambda} \, ,
\end{align}
we can choose $\Lambda^{(M)} = \overbar{\Lambda}$ and $\Lambda^{(E)} = \overbar{\Lambda}$, and this also gives a solution to (\ref{maxwell_ansatz}).

In the flat space case, one can explicitly construct infinitely many one-forms with either self-dual or anti-self-dual derivatives, which therefore lead to infinitely many conserved charges. One way to see this is by solving the self-duality constraint in twistor space. Suppose that we convert Lorentz indices to pairs of spinor indices as
\begin{align}
    \Lambda_{\alpha \dot{\alpha}} = \sigma^\mu_{\alpha \dot{\alpha}} \Lambda_\mu \, ,
\end{align}
where the $\sigma^\mu_{\alpha \dot{\alpha}}$ are Pauli matrices. Given a spinor $\lambda_{\dot{\alpha}}$, define
\begin{align}
    z^\alpha = x^{\alpha \dot{\alpha}} \lambda_{\dot{\alpha}} \, , 
\end{align}
where $x^{\alpha \dot{\alpha}}$ are the bispinor versions of the spacetime coordinates $x^\mu$. Furthermore, given such a $\lambda_{\dot{\alpha}}$, let $\mu_{\dot{\alpha}}$ be an arbitrary reference spinor which is not collinear with $\lambda_{\dot{\alpha}}$. Then consider the one-form $\Lambda_{\alpha \dot{\alpha}}$ defined by the twistor transform
\begin{align}\label{sd_lambda_twistor}
    \Lambda_{\alpha \dot{\alpha}} = \frac{1}{2 \pi i} \oint_{\mathcal{C}} \, d \lambda^{\dot{\gamma}} \, \lambda_{\dot{\gamma}} \,  \frac{\mu_{\dot{\alpha}}}{\epsilon^{\dot{\delta} \dot{\epsilon}} \mu_{\dot{\delta}} \lambda_{\dot{\epsilon}}} \frac{\partial}{\partial z^\alpha} \varphi ( z , \lambda ) \, ,
\end{align}
where the integral is taken over a closed curve $\mathcal{C}$ in $\mathbb{CP}^1$, and $\varphi$ is a holomorphic function on $\mathbb{CP}^3$, which is coordinatized by the twistor variables $z^\alpha$ and $\lambda_{\dot{\alpha}}$. The measure $d \lambda^{\dot{\gamma}} \, \lambda_{\dot{\gamma}}$ on $\mathbb{CP}^1$ has projective weight $+2$ and thus the remainder of the integrand must be homogeneous with degree $-2$ in order for (\ref{sd_lambda_twistor}) to be well-defined. This integral is then independent of the choice of $\mu_{\dot{\alpha}}$ and of the precise contour of integration, and it implies that the exterior derivative of $\Lambda$ can be written as
\begin{align}
    \left( d \Lambda \right)_{\alpha \dot{\alpha} \beta \dot{\beta}} = \left( d \Lambda \right)_{\alpha \beta} \epsilon_{\dot{\alpha} \dot{\beta}} \, ,
\end{align}
which guarantees that $d \Lambda$ is self-dual and its self-dual part is encoded in the object $\left( d \Lambda \right)_{\alpha \beta}$. A similar twistor construction can be used to solve the constraint for a one-form with anti-self-dual exterior derivative. For an introduction to twistor techniques in field theory, see the lecture notes \cite{Wolf:2010av,Adamo:2017qyl} or the textbook \cite{Huggett:1986fs}.

One may then conclude that, as a consequence of the fact that there are infinitely many holomorphic functions $\varphi ( z, \lambda )$ on twistor space, we can likewise find infinitely many one-forms $\Lambda$ whose derivatives satisfy either a self-duality or anti-self-duality condition. By the argument above, this leads to infinitely many solutions of (\ref{maxwell_ansatz}) and thus infinitely many non-trivial conserved currents in the Maxwell theory. It was emphasized in \cite{Hofman:2018lfz} (see also \cite{Hofman:2024oze}) that the presentation of solutions to the self-duality conditions in twistor space makes the analogy between the conserved currents in the Maxwell theory and those in a $2d$ CFT more transparent; in that work the algebra of the non-trivial charges was also computed.

To conclude this subsection, let us give another interpretation of this infinite collection of conserved currents. We have explained above that the currents (\ref{maxwell_ansatz}) may be viewed as being in one-to-one correspondence with either (i) harmonic two-forms, or (ii) pairs of one-forms $\Lambda$, $\overbar{\Lambda}$ whose exterior derivatives are self-dual and anti-self-dual. A more physical view of this correspondence principle is as follows. Consider a given, fixed field configuration $\widehat{F}_{\mu \nu}$ which solves the vacuum Maxwell equations,
\begin{align}\label{maxwell_solns}
    d \widehat{F} = d \ast \widehat{F} = 0 \, .
\end{align}
This means that $\widehat{F}$ is a harmonic $2$-form, and can be locally trivialized as $\widehat{F} = d \widehat{A}$, $\ast \widehat{F} = d \widehat{B}$. Therefore, choosing the parameters
\begin{align}\label{maxwell_soln_one_forms}
    \Lambda^{(E)} = \widehat{A} \, , \qquad \Lambda^{(M)} = - \widehat{B} \, ,
\end{align}
furnishes us with a solution to the constraint (\ref{maxwell_vanish_intermediate}). Given such a solution, we may choose to ``forget'' that these one-forms (\ref{maxwell_soln_one_forms}) have an interpretation in terms of solutions to the Maxwell equations, and simply treat them as abstract forms which can be used to construct a current $j^\mu$ as in (\ref{maxwell_ansatz}). The resulting $j^\mu$ is then conserved.

Therefore, a third way of interpreting the argument of this subsection is as follows. For each fixed solution $\widehat{F}_{\mu \nu}$ to the equations of motion for the free Maxwell theory, we can define an associated current $j^\mu$ using equations (\ref{maxwell_soln_one_forms}) and (\ref{maxwell_ansatz}). This current is then conserved, by virtue of equation (\ref{current_cons_lambdas_condition}). In particular, we emphasize that conservation of this current holds whenever the equations of motion are satisfied. This means that $\partial_\mu j^\mu = 0$ for any on-shell field configuration $F_{\mu \nu}$, not only for the fixed solution $\widehat{F}_{\mu \nu}$ which was used for its construction. 

Said differently, since we have seen that every harmonic $2$-form gives rise to a conserved current -- and because every $\widehat{F}_{\mu \nu}$ which satisfies (\ref{maxwell_solns}) is a harmonic $2$-form -- we obtain one such current $j^\mu$ for each solution of the Maxwell equations.

\subsection{Generalization to Interacting Theories}\label{sec:generalization_interacting}

Because Maxwell theory is a free model, and all free models possesss infinitely many conserved quantities, it is perhaps unsurprising that one can build infinitely many currents from the higher-form global symmetries in the manner described here. It would be more interesting if such an argument could be used to find conserved currents in interacting theories, such as models of non-linear electrodynamics, perhaps with an additional assumption such as duality invariance. We now turn to the question of this generalization.

In a general theory of non-linear electrodynamics, the appropriate modification of the ansatz (\ref{maxwell_ansatz}) for a conserved current is
\begin{align}\label{NLED_ansatz}
    j^\mu = \widetilde{G}^{\mu \nu} \Lambda_\nu^{(E)} + \widetilde{F}^{\mu \nu} \Lambda_\nu^{(M)} \, , 
\end{align}
because the equations of motion and Bianchi identity guarantee the conservation of $\widetilde{G}^{\mu \nu}$ and $\widetilde{F}^{\mu \nu}$, respectively. The divergence of this would-be current is
\begin{align}\label{NLED_current_condition}
    \partial_\mu j^\mu = \widetilde{G}^{\mu \nu} \partial_\mu \Lambda_\nu^{(E)} + \widetilde{F}^{\mu \nu} \partial_\mu \Lambda_\nu^{(M)} \, ,
\end{align}
which we would like to vanish identically.

Solving the condition (\ref{NLED_current_condition}) in a general theory of non-linear electrodynamics seems difficult. We will therefore restrict to duality-invariant theories. Given this additional assumption, let us first point out that there is at least one simple solution to the condition (\ref{NLED_current_condition}) which one can immediately write down. Because
\begin{align}
    d F = d G = 0 \, ,
\end{align}
one can trivialize both $2$-forms as
\begin{align}
    F = d A \, , \qquad G = d B \, ,
\end{align}
and then choosing
\begin{align}\label{simple_solution}
    \Lambda_\mu^{(E)} = B_\mu \, , \qquad \Lambda_\mu^{(M)} = A_\mu \, ,
\end{align}
the divergence of $j^\mu$ is
\begin{align}\label{BB_current}
    \partial_\mu j^\mu = \widetilde{G}^{\mu \nu} G_{\mu \nu} + \widetilde{F}^{\mu \nu} F_{\mu \nu} \, ,
\end{align}
which vanishes precisely due to the duality-invariance condition (\ref{BB_condition}). The conservation of this current was already noticed in the early work of \cite{Gaillard:1981rj}. However, the present discussion now allows us to re-interpret this current as one which descends from the $1$-form global symmetries in a general theory of duality-invariant electrodynamics, in exactly the same way that the currents of \cite{Hofman:2018lfz} arise in Maxwell theory.

Let us now ask whether there are other solutions to (\ref{NLED_current_condition}) besides the simple one (\ref{simple_solution}). To address this question, we will use the result of section \ref{sec:duality_and_metric} that the equations of motion for any duality-invariant theory can be written as
\begin{align}
    \partial_\mu \left( \left( h^{-1} \right)^{\mu \rho} F_{\rho \sigma} \left( h^{-1} \right)^{\sigma \nu} \right) = 0 \, ,
\end{align}
for a field-dependent metric $h_{\mu \nu}$ with inverse $\left( h^{-1} \right)^{\mu \nu}$. The condition (\ref{NLED_current_condition}) for conservation of the candidate current then reads
\begin{align}\label{duality_invariant_current_condition}
    \partial_\mu j^\mu = \left( \left( h^{-1} \right)^{\mu \rho} F_{\rho \sigma} \left( h^{-1} \right)^{\sigma \nu} \right) \partial_\mu \Lambda_\nu^{(E)} + \widetilde{F}^{\mu \nu} \partial_\mu \Lambda_\nu^{(M)} = 0 \, .
\end{align}
We recall that indices are still raised or lowered with the background metric $g_{\mu \nu}$. It is useful to define the Hodge star operation with respect to the field-dependent metric $h_{\mu \nu}$, which acts on a $2$-form $A_{\mu \nu}$ as
\begin{align}\label{h_hodge_star}
    \left( \ast_h \; A \right)_{\nu_3 \nu_4} = \frac{\sqrt{ | \det ( h ) | }}{2!} A_{\mu_1 \mu_2} \left( h^{-1} \right)^{\mu_1 \nu_1} \left( h^{-1} \right)^{\mu_2 \nu_2} \epsilon_{\nu_1 \nu_2 \nu_3 \nu_4} \, .
\end{align}
As we showed above, $| \det ( h ) | = 1$ for the field-dependent metric associated with any theory of duality-invariant electrodynamics, so we drop this factor. Then in terms of the operation (\ref{h_hodge_star}), the condition (\ref{duality_invariant_current_condition}) is equivalent to
\begin{align}
    F \wedge \left( d \Lambda^{(M)} + \ast_h d \Lambda^{(E)}  \right) = 0 \, ,
\end{align}
which one can solve by taking
\begin{align}\label{duality_invariant_twisted_self_duality}
    d \Lambda^{(M)} = - \ast_h  d \Lambda^{(E)} \, ,
\end{align}
where we have used that $\ast_h \ast_h = 1$ in Euclidean signature. We see that the condition (\ref{duality_invariant_twisted_self_duality}) straightforwardly generalizes the constraint relating $\Lambda^{(E)}$ and $\Lambda^{(M)}$ in the Maxwell case, equation (\ref{harmonic_solution}), to which (\ref{duality_invariant_twisted_self_duality}) reduces in the case where $h_{\mu \nu} = g_{\mu \nu}$.

The condition (\ref{duality_invariant_twisted_self_duality}) implies that both $d \Lambda^{(E)}$ and $d \Lambda^{(M)}$ are harmonic $2$-forms with respect to the field-dependent metric $h_{\mu \nu}$:
\begin{align}\label{geometrical_characterization}
    0 = d \left( d \Lambda^{(E)} \right) = d \left( \ast_h d \Lambda^{(E)} \right) \, , \qquad 0 = d \left( d \Lambda^{(M)} \right) = d \left( \ast_h d \Lambda^{(M)} \right) \, .
\end{align}
One can therefore build a conserved current for each one-form $\Lambda^{(M)}$ whose derivative $d \Lambda^{(M)}$ is  harmonic with respect to $h_{\mu \nu}$. Given such a $\Lambda^{(M)}$, we define $d \Lambda^{(E)}$ by (\ref{duality_invariant_twisted_self_duality}), which is closed because $d \Lambda^{(M)}$ is co-closed, and can therefore be locally trivialized to a one-form $\Lambda^{(E)}$. We say that there is one current for each ``$h$-harmonic form'' on the spacetime manifold $\mathcal{M}$. Alternatively, one can perform a splitting of $d \Lambda^{(E)}$ and $d \Lambda^{(M)}$ into self-dual and anti-self-dual parts, where duality is defined with respect to $\ast_h$. It follows that we can construct one non-trivial current for each one-form $\Lambda$ with $h$-self-dual or $h$-anti-self-dual derivative $d \Lambda$.

We pointed out, around equation (\ref{maxwell_solns}), that the infinite set of conserved currents in Maxwell theory can also be described as being in 1-to-1 correspondence with solutions to the equations of motion. This is not quite true for the generalization we have just discussed, for the following reason. Consider a theory of non-linear electrodynamics whose equations of motion are equivalent to those of Maxwell theory on the field-dependent metric $h_{\mu \nu} ( F ) $. Choose a fixed field configuration $\widehat{F}_{\mu \nu}$, and let $\hat{h}_{\mu \nu} = h_{\mu \nu} ( \widehat{F} )$ be the metric evaluated using the configuration $\widehat{F}_{\mu \nu}$ rather than the abstract variable $F_{\mu \nu}$. Then $\widehat{F}_{\mu \nu}$ solves the equations of motion if
\begin{align}
    d \widehat{F} = d \ast_{\hat{h}} \widehat{F} = 0 \, ,
\end{align}
whereas the parameter $\Lambda$ which generates a conserved current in the theory obeys
\begin{align}
    d ( d \Lambda ) = d ( \ast_{h} d \Lambda ) = 0 \, .
\end{align}
The difference between these conditions is that one uses the Hodge star associated with $h_{\mu \nu} ( F )$ and the other uses the Hodge star built from $\hat{h}_{\mu \nu} = h_{\mu \nu} ( \widehat{F} )$. It is therefore not immediate that conserved currents in a general duality-invariant theory can be put into correspondence with solutions to the equations of motion.

Nonetheless, the correspondence between conserved currents and $h$-harmonic two-forms implied by (\ref{geometrical_characterization}) still holds in any duality-invariant theory. We have therefore geometrized the characterization of conserved currrents which descend from the $1$-form global symmetries in any such self-dual theory. 

\subsection{Remarks on Generalized Currents}

We conclude this section with a few further comments regarding the geometrical description of conserved currents in theories of duality-invariant electrodynamics.

\subsubsection*{\ul{\it Compact Riemannian case}}

Suppose that we are studying the Maxwell theory on a compact Riemannian spacetime manifold $\mathcal{M}$ (i.e. with Euclidean signature). In this case, the space of harmonic $p$-forms on a manifold $\mathcal{M}$ is isomorphic to the $p$-th de Rham cohomology group, $H^p ( \mathcal{M} )$. Although the definition of a harmonic form relies on the details of the Riemannian metric, importantly, the de Rham cohomology is defined \emph{independently of the metric}. That is, the de Rham cohomology is a topological quantity which depends only on the smooth structure of $\mathcal{M}$ but not on its metric structure.

Therefore, at least for two conventional field-independent Riemannian metrics $g_{\mu \nu}$ and $h_{\mu \nu}$, given any $2$-form which is harmonic on $\mathcal{M}$ with respect to $g_{\mu \nu}$, there is a corresponding $2$-form which is harmonic on $\mathcal{M}$ with respect to $h_{\mu \nu}$. Pictorially,
\begin{align}\label{isomorphisms}
    \mathrm{Harm}^2 ( \mathcal{M} , g ) \cong H^2 ( \mathcal{M} ) \cong \mathrm{Harm}^2 ( \mathcal{M} , h ) \, ,
\end{align}
where $\mathrm{Harm}^p ( \mathcal{M} , g )$ denotes the space of harmonic $p$-forms on a manifold $\mathcal{M}$ with metric $g_{\mu \nu}$. Assuming that a similar isomorphism continues to hold for the case of field-dependent metrics $h_{\mu \nu}$, there would be a bijection between $g$-harmonic and $h$-harmonic two-forms. In particular, the isomorphisms (\ref{isomorphisms}) would then suggest that, for any theory of duality-invariant non-linear electrodynamics defined on a compact Riemannian manifold, there exist as many conserved currents $j^\mu$ which arise from the solutions to the condition (\ref{NLED_current_condition}) as there are corresponding conserved currents for the free Maxwell theory defined on the same spacetime manifold.

\subsubsection*{\ul{\it Analogy with $\TT$-deformed $2d$ conformal field theories}}

The infinite dimensional algebra of charges in the Maxwell CFT has many properties in common with the symmetry algebra of a $2d$ CFT. Because an irrelevant deformation, such as the $4d$ $\TT$ flow which deforms the Maxwell theory into the Born-Infeld model, would appear to destroy the structure of such an infinite-dimensional algebra, one might be skeptical that the Maxwell currents indeed have analogues in other theories of duality-invariant electrodynamics. Equivalently, one might doubt whether any $h$-harmonic forms, which generate these charges, actually exist.

Let us provide some circumstantial evidence that such solutions might be expected to exist, by analogy with $\TT$ deformations of CFTs in $2d$. In this setting, the undeformed CFT possesses a $\mathrm{Virasoro} \times \mathrm{Virasoro}$ symmetry, which likewise appears to be broken by the irrelevant $\TT$ flow, as it destroys conformal symmetry and introduces a scale $\lambda$. However, despite this expectation, a $\TT$-deformed CFT still possesses a full $\mathrm{Virasoro} \times \mathrm{Virasoro}$ symmetry at the quantum level \cite{Guica:2021pzy}. The generators of this deformed symmetry algebra can be understood via a field-dependent diffeomorphism, which means that the new symmetry generators are ``dressed'' with field-dependent factors compared to the Virasoro generators in the seed CFT.

These field-dependent conformal symmetries, or ``pseudo-conformal symmetries,'' have also been investigated at the classical level \cite{Guica:2020uhm,Guica:2022gts}. The classical structure of pseudo-conformal symmetry generators is most relevant for our present purposes, since we have restricted ourselves to an essentially classical analysis at the level of $4d$ effective field theory. Because the behavior of the infinite charge algebra in the $4d$ Maxwell theory mirrors that of a $2d$ CFT -- and since an irrelevant deformation of a $2d$ CFT, namely $\TT$, is known to possess a deformed infinite charge algebra with field-dependent generators -- it stands to reason that similar deformed charges may exist for theories of $4d$ electrodynamics which satisfy stress tensor flow equations. Indeed, the use of our field-dependent metric $h_{\mu \nu}$ appears to enact a field-dependent dressing of the symmetry generators precisely as in the $2d$ case.

\subsubsection*{\ul{\it The self-dual sector of theories of electrodynamics}}

The vacuum Maxwell equations, $d F = d \ast F = 0$, are automatically satisfied by any (anti-)self-dual form, $F = \pm \ast F$, since $F = d A$ is closed. The collection of all such solutions has sometimes been called the ``(anti-)self-dual sector'' of the Maxwell theory \cite{self_dual}; this sector has real solutions for the field strength in Euclidean signature, but complex field strengths for Lorentz signature. In the non-Abelian setting, the analogous (anti-)self-dual sector of Yang-Mills theory has been extensively studied (where such configurations are referred to as instanton solutions), and has deep connections to integrability.\footnote{See, for instance, \cite{Ablowitz:1993ec,Mason:1991rf,Bardeen:1995gk} and references therein for further details.} Note that, for any gauge group $G$, the energy-momentum tensor $T_{\mu \nu}$ vanishes on any self-dual or anti-self-dual solution of the Yang-Mills equations, including in the Abelian case $G = U ( 1 )$. 

In a more general theory of non-linear electrodynamics, one may likewise attempt to solve the equations of motion $d F = d G = 0$ by setting $F = \pm G$. Within a duality-invariant theory, which can be realized with a field-dependent metric, this condition can be stated geometrically: in such cases one seeks to impose
\begin{align}
    \widetilde{G}^{\mu \nu} = \left( h^{-1} \right)^{\mu \rho} F_{\rho \sigma} \left( h^{-1} \right)^{\sigma \nu} = \pm \widetilde{F}^{\mu \nu} \, ,
\end{align}
or equivalently,
\begin{align}\label{sd_sector}
    F = \pm \ast_h F \, .
\end{align}
We will likewise refer to solutions of the equation (\ref{sd_sector}), assuming they exist, as the (anti-)self-dual sector of any theory of duality-invariant electrodynamics. 
In this sector, the construction of the conserved currents outlined in section \ref{sec:generalization_interacting} simplifies. Here the condition on the parameters $\Lambda^{(E)}$, $\Lambda^{(M)}$ is simply $d \Lambda^{(M)} = - \ast d \Lambda^{(E)}$, which is the same as the condition (\ref{harmonic_solution}) in the Maxwell theory. This suggests that, within any duality-invariant model, there may exist a sector of solutions to the equations of motion which is identical to the corresponding sector of solutions for the Maxwell theory, and for which the infinite charge algebra has identical properties.

The results of this work provide a heuristic explanation for why this might be the case. We have already mentioned that duality-invariant theories can be obtained by deforming the Maxwell theory by functions of the energy-momentum tensor. For any (anti-)self-dual solution of the Maxwell equations, the stress tensor vanishes, so one expects that this sector of solutions is unaffected by any deformation by $T_{\mu \nu}$. This agrees with the expectation that the symmetry structure of the (anti-)self-dual sector of a duality-invariant theory is identical to that of the free Maxwell theory.

\subsubsection*{\ul{\it Lipkin's zilch}}

We argued above that, if $h$-harmonic forms may be constructed from $g$-harmonic forms, then a general duality-invariant theory of electrodynamics may have a large collection of conserved quantities, which descend from the $1$-form global symmetry structure in the model. At least in the Maxwell theory, this infinite set of conserved quantities can be explicitly constructed. One might even be led to conjecture that all of the conserved quantities can be interpreted in terms of higher-form global symmetries in this way, at least for the free Maxwell case. We now provide a counter-example to this conjecture, which illustrates that not all of the conserved quantities in such a theory can be straightforwardly understood via higher-form symmetries.

An interesting conserved quantity which is not obviously related to the structure we have discussed above is called Lipkin's zilch \cite{lipkin}. In the Maxwell theory on flat Minkowski space, zilch is defined by
\begin{align}
    Z_{\mu \nu \rho} = \tensor{\widetilde{F}}{_\mu^\lambda} \partial_\rho F_{\lambda \nu} - \tensor{F}{_\mu^\lambda} \partial_\rho \widetilde{F}_{\lambda \nu} \, .
\end{align}
Using the conservation of the $2$-form currents $\widetilde{F}^{\mu \nu}$ and $F^{\mu \nu}$, one can show that the divergence of the zilch tensor with respect to any of its indices vanishes:
\begin{align}
    \partial^\mu Z_{\mu \nu \rho} = \partial^\nu Z_{\mu \nu \rho} = \partial^\rho Z_{\mu \nu \rho} = 0 \, .
\end{align}
Furthermore, the zilch tensor enjoys the tracelessness conditions $\tensor{Z}{^\mu_\mu_\nu} = \tensor{Z}{^\mu_\nu_\mu} = 0$. See \cite{kibble} for details on the proofs of these properties and for equivalent rewritings of the zilch tensor. One component of the zilch tensor is related to a quantity known as optical chirality, which was introduced and shown to be conserved in \cite{optical_chirality}.

However, the zilch tensor is not associated with a $2$-form global symmetry, which would give rise to a conserved $3$-form current $J_{\mu \nu \rho}$. The reason for this is simply that the zilch tensor does not have the correct symmetry properties on its indices; whereas a $3$-form current $J_{\mu \nu \rho}$ must be totally antisymmetric, the zilch tensor is symmetric on its first two indices, $Z_{\mu \nu \rho} = Z_{\nu \mu \rho}$. Although it is not connected to generalized global symmetries, the conservation of the zilch tensor has been understood via a symmetry in a duality-invariant presentation of Maxwell theory \cite{Aghapour:2019iav}, and in terms of a hidden symmetry algebra in the standard formulation \cite{Letsios:2022bid}.

It would be interesting to investigate whether the field-dependent metric perspective could be used to define a more general notion of the zilch tensor which is conserved in any theory of duality-invariant electrodynamics, perhaps by forming a combination like $Z_{\mu \nu \rho} = \tensor{\widetilde{F}}{_\mu^\lambda} \partial_\rho \widetilde{G}_{\lambda \nu} - \tensor{\widetilde{G}}{_\mu^\lambda} \partial_\rho \widetilde{F}_{\lambda \nu}$ and using conservation of $\widetilde{G}$, $\widetilde{F}$.

\section{Conclusion}\label{sec:conclusion}

In this work, we have geometrized the space of theories of duality-invariant electrodynamics through the introduction of a field-dependent metric. That is, we established an equivalence between the condition of electric-magnetic duality invariance and the statement that the equations of motion for a theory can be recast as the Maxwell equations on a unit-determinant field-dependent metric $h_{\mu \nu} ( F )$.

We then used this geometrical perspective to offer additional insight into the symmetry structure of duality-invariant theories. For instance, we have argued that the electric and magnetic $1$-form global symmetries in a model of self-dual electrodynamics can be used to construct one conserved current for each $h$-harmonic $2$-form, which recovers the result of \cite{Hofman:2018lfz,Hofman:2024oze} in the case of Maxwell theory.

There remain several interesting directions for future research. One natural question is whether, given a field-dependent metric $h_{\mu \nu}$ of the type considered here, one can formally construct $h$-harmonic $2$-forms from $g$-harmonic $2$-forms. This would roughly correspond to a map between solutions to the equations of motion for Maxwell theory and solutions to the equations of motion for a general duality-invariant theory. Although the existence of such a map may seem rather unlikely, we should point out that similar dictionaries exist for stress tensor deformations in other settings. For instance, in the case of the $2d$ $\TT$ deformation, \cite{Conti:2018tca} showed that solutions of the deformed equations of motion can be generated from solutions of the undeformed equations of motion by a field-dependent diffeomorphism. It would be interesting to see whether a similar procedure exists for general stress tensor deformations in $4d$, which would give a method for building $h$-harmonic forms.

Another direction is to investigate $6d$ analogues of the structures considered in this work. There are many similarities between duality-invariant theories of $4d$ electrodynamics and chiral tensor theories in six spacetime dimensions. For instance, as we alluded to above, stress tensor deformations of $4d$ duality-invariant theories preserve duality invariance, whereas deformations of $6d$ chiral tensor theories in the PST formulation preserve the condition of PST gauge invariance \cite{Ferko:2024zth}. Motivated by these analogies, it would be interesting to see if a general $6d$ chiral tensor theory can be represented as a free chiral tensor theory on a field-dependent metric.\footnote{This may also aid in the study of the $6d$ ModMax-like chiral tensor theory \cite{Bandos:2020hgy}; see \cite{Deger:2024jfh} for recent progress on the construction of solutions to this theory coupled to gravity.}

We hope to return to these questions in future work.

\begin{acknowledgement}

We are especially grateful to Dmitri P. Sorokin for many useful discussions and collaboration in the early stages of this project. We also thank Diego Hofman and Gabriele Tartaglino-Mazzucchelli for helpful comments, and we thank T. Daniel Brennan, Tommaso Morone, and an anonymous referee for feedback on a draft of this article. CF and CLM acknowledge kind hospitality and financial support at the MATRIX Program ``New Deformations of Quantum Field and Gravity Theories'' and thank the participants of this meeting for productive conversations on related topics. CF is supported by U.S. Department of Energy grant DESC0009999 and by funds from the University of California. CLM is supported by a postgraduate scholarship at the University of Queensland and was partially supported by the Australian Research Council (ARC) Future Fellowship FT180100353, and a Capacity Building Package of the University of Queensland.
\end{acknowledgement}

\bibliographystyle{utphys}
\bibliography{master}

\end{document}